\documentclass[onecolumn,preprintnumbers,amsmath,amssymb,floatfix,11pt]{revtex4}
\usepackage{graphicx}
\usepackage{dcolumn}
\usepackage{bm}
\usepackage{graphics}
\usepackage{amssymb}
\usepackage{amsmath}
\usepackage{xspace}
\usepackage{epsfig}
\newcommand {\etal}{\begin{itshape}et al\end{itshape}. }

\newcommand {\ET}{ET }

\newcommand {\bI}{$\beta$-ET$_2$I$_3$ }
\newcommand {\bIc}{$\beta$-ET$_2$I$_3$, }
\newcommand {\bHI}{$\beta_\textrm{H}$-ET$_2$I$_3$ }
\newcommand {\bLI}{$\beta_\textrm{L}$-ET$_2$I$_3$ }
\newcommand {\bH}{$\beta_\textrm{H}$ }
\newcommand {\bL}{$\beta_\textrm{L}$ }

\newcommand {\Br}{$\kappa$-ET$_2$Cu[N(CN)$_2$]Br }

\begin{document}

\title{The origin of the difference in the superconducting critical
temperatures of the \bH and \bL phases of (BEDT-TTF)$_2$I$_3$}
\author{B. J. Powell\footnote{Electronic address: powell@physics.uq.edu.au}}
\address{Department of Physics, University of Queensland, Brisbane,
Queensland 4072, Australia}




\begin{abstract}
Incommensurate lattice fluctuations are present in
the \bL phase ($T_c\sim1.5$~K) of ET$_2$I$_3$ (where ET is
BEDT-TTF - bis(ethylenedithio)tetrathiafulvalene) but are absent
in the \bH phase ($T_c\sim7$~K). We propose that the disorder in
the conformational degrees of freedom of the terminal ethylene
groups of the ET molecules, which is required to stabilise the
lattice fluctuations, increases the quasiparticle scattering rate
and that this leads to the observed difference in the
superconducting critical temperatures, $T_c$, of the two phases.
We calculate the dependence of $T_c$ on the interlayer residual
resistivity. Our theory has no free parameters. Our predictions
are shown to be consistent with experiment. We describe
experiments to
conclusively test our hypothesis.
\end{abstract}

\maketitle

\vspace{10pt}

\noindent It has long been known that the application of
hydrostatic pressure, $P$, to \bI has a dramatic effect on the
superconducting critical temperature, $T_c$. At ambient pressure
$T_c \sim 1.5$~K but when the applied pressure reaches $P\sim
1$~kbar a discontinuous increase in $T_c$ ($\sim 7$~K) is
observed. The low $T_c$ state ($P \lesssim 1$~kbar) is denoted the
\bL phase and the high $T_c$ state ($P \gtrsim 1$~kbar) is
labelled the \bH phase. When the pressure on the \bH phase is
decreased the material does not return to the \bL phase but rather
$T_c$ is seen to further increase. Below $T\sim 130$~K the
resistivity of the \bH phase is found to undergo a discontinuous
decrease while no such anomaly is found in the \bL phase
\cite{Ginodman}. Incommensurate lattice fluctuations have been
observed in the \bL phase but they are absent in the \bH phase
below $T\sim130$~K \cite{Emge}. The incommensurate lattice
fluctuations are stabilised by variations in the conformational
ordering of the terminal ethylene groups of the \ET molecules and
thus can only exist in the presence of disorder \cite{Ravy}. For a
recent review of this phenomenology see \cite{Ishiguro}.

Although the change in resistivity between the \bL and \bH phases
and the incommensurate lattice fluctuations have been studied both
theoretically \cite{Ravy} and experimentally
\cite{Ginodman,Emge,Ishiguro}, no explanation of the change in
$T_c$ has been forthcoming. In this paper we propose that
difference in the $T_c$'s of the two phases is due to disorder, we
will show that current data is consistent with our explanation and
suggest experiments which could clearly determine whether or not
this is the correct explanation of the difference in $T_c$.

In a recent paper it was shown \cite{BenRHMdisorder} that
intrinsically non-magnetic disorder decreases the $T_c$ of a wide
range of the superconducting salts of \ET in line with the
Abrikosov--Gorkov (AG) formula:
\begin{eqnarray}
\ln \left(\frac{T_{c0}}{T_{c}} \right) = \psi\left( \frac{1}{2} +
\frac{\hbar}{4\pi k_BT_{c}}\frac{1}{\tau} \right) - \psi\left(
\frac{1}{2} \right), \label{eqn:AG}
\end{eqnarray}
where $T_{c0}$ is the superconducting transition temperature of a
pure sample, $1/\tau$ is the quasiparticle scattering rate and
$\psi(x)$ is the digamma function. There are two scenarios
compatible with this observation: either there is opposite spin
pairing and the impurities induce localised magnetic moments, or
else there is a finite angular momentum pairing state (most
probably d-wave pairing). In this paper we will not discuss which
of these scenarios is realised, but rather make use of this
observation of the effect that increasing $1/\tau$ has on $T_c$.

In a quasi-two dimensional metal, such as \bIc the residual
interlayer resistivity is given by
\begin{eqnarray}
\rho_0 = \frac{\pi\hbar^4}{2e^2m^*ct_\perp^2}\frac{1}{\tau}
\label{eqn:interlayer_resistivity}
\end{eqnarray}
where $m^*$ is the quasiparticle effective mass, $c$
($=15.291$~\AA~for \bI\cite{Ishiguro}) is the interlayer lattice
constant and $t_\perp$ is interlayer hopping integral.
Substituting (\ref{eqn:interlayer_resistivity}) into
(\ref{eqn:AG}) one finds that
\begin{eqnarray}
\ln \left(\frac{T_{c0}}{T_{c}} \right) = \psi\left( \frac{1}{2} +
\frac{e^2m^*ct_\perp^2}{2\pi^2\hbar^3k_BT_{c}}\rho_0 \right) -
\psi\left( \frac{1}{2} \right). \label{eqn:powell}
\end{eqnarray}
from quantum oscillation Wosnitza \etal \cite{Wosnitza_ratio_kIBr}
found that $m^*/m_e=4.2\pm 0.2$, where $m_e$ is the electronic
rest mass, $t_\perp/E_F = 1/(175\pm 10)$, where $E_F$ is the Fermi
energy, and the Fermi wavevector, $k_F = (3.4 \pm 0.2) \times
10^9$~cm$^{-1}$. Thus taking $E_F = \hbar^2 k_F^2/2m^*$ one finds
that $t_\perp = 0.59 \pm 0.08$~meV. $T_{c0}=7.75$~K is found from
fast electron irradiation experiments \cite{BenRHMdisorder,Forro}.

In figure 1 we plot $T_c$ as a function of $\rho_0$. We stress
that there are no free parameters in this plot once the data from
Wosnitza \etal and Forro \etal\cite{BenRHMdisorder,Forro} is
considered. For comparison we also plot the data of Tokumoto \etal
\cite{Tokumoto} who deliberately induced impurities in \bHI by
fabricating the alloy
$\beta$-ET$_2$(I$_3$)$_{1-x}$(IBr$_2$)$_x$. The data is consistent
with our theory although clearly more data is required to properly
test our prediction.


\begin{figure}
    \centering
    \epsfig{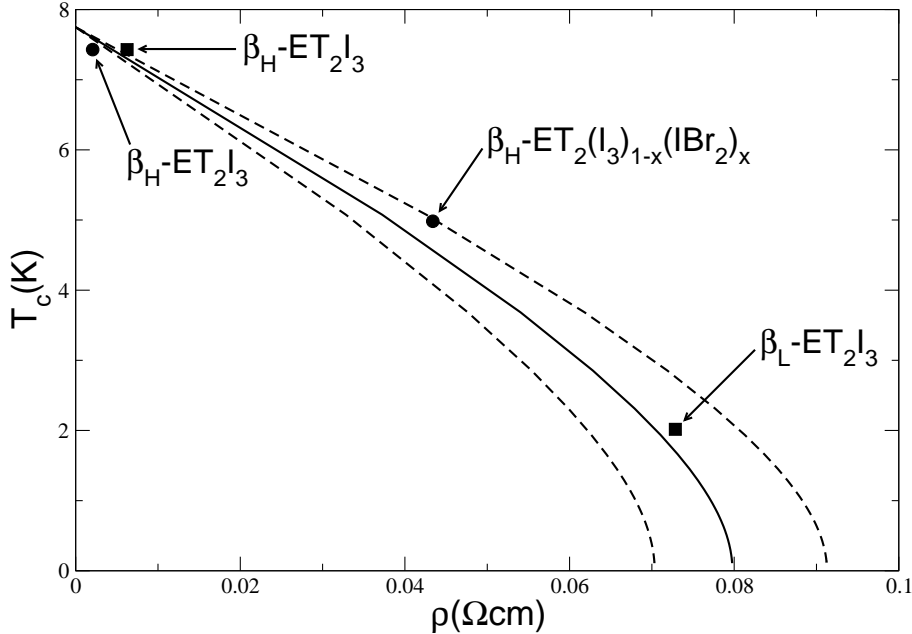} \label{figure}
\caption{The
 variation of the superconducting transition
    temperature of \bHI with the interlayer resistivity. The prediction (solid line) is
    made from equation (\ref{eqn:powell}) taking the measured interlayer hopping
    integral, $t_\perp$
    \cite{Wosnitza_ratio_kIBr}. The dotted lines
    show the effect of changing $t_\perp$ by one standard deviation. The critical
    temperature of a pure sample, $T_{c0}$, is that found from fast electron irradiation
    experiments \cite{BenRHMdisorder,Forro}, we do not include the effects of the errors in this
    measurement. By way of comparison we show data
    for impurities induced in \bHI by fabricating the alloy
    $\beta$-ET$_2$(I$_3$)$_{1-x}$(IBr$_2$)$_x$
    \cite{Tokumoto} (circles). This data is consistent with our prediction.
    The squares are the data \cite{Ginodman} for both the \bH and \bL
    phases. However, as the size of the samples was not reported the value of
    the resistivity contains one free parameter, the relative dimension of the sample,
    $A/d$. However,
this data does show that the relative change
    in resistivity between the \bL and \bH phases is consistent with our hypothesis that
    the disorder required to stabilise the incommensurate lattice fluctuations
    observed in the \bL phase lowers $T_c$ by increasing the quasiparticle
    scattering rate, $1/\tau$.}
\end{figure}


We propose that the disorder in the conformational degrees of
freedom of the terminal ethylene groups of the ET molecules which
is required to stabilise incommensurate lattice fluctuations in
the \bL phase \cite{Ravy}, but absent in the \bH phase, increases
the quasiparticle lifetime. (Either by inducing localised magnetic
moments or causing variation in the site energy, depending on
which of the scenarios proposed in \cite{BenRHMdisorder} is
realised). Note that conformational disorder of the terminal
ethylene groups of \Br can be controlled by varying the cooling
rate and this leads to variations in $T_c$ which are well
described by the AG formula \cite{BenRHMdisorder}.

Unfortunately, we are not aware of any reports of the resistivity
of \bLI with which to test our hypothesis. For example Ginodman
\etal \cite{Ginodman} reported the change in resistance observed
between the \bH and \bL phases, but did not report the size of
their crystals. Taking the relative dimensions of their sample as
a fitting parameter we find good agreement with our prediction
(see figure 1; based on $A/d = 0.4$~cm, where $A$ is the
cross-sectional area of the sample and $d$ is the length of the
sample in the c-axis, this is actually a fairly typical relative
dimension for samples of this material). However, as we do not
know the size of the sample this does not represent a very
stringent test of our hypothesis. The resistance data of Ginodman
\etal is well described by the Fermi liquid form $R(T) = R_0 +
AT^2$ in both the \bL and \bH phases with the same value of $A$ in
both phases. This suggests that major difference between the two
phases is the scattering rate due to impurities, in support of our
hypothesis.

Systematic measurements of the resistivity of \bI in both the \bL
and \bH phases are required to test our hypothesis. If this were
done for several samples with varying amounts of disorder due to
random imperfections produced during the fabrication process then
it would be possible to map out the entire curve of figure 1 for
the \bH phase. Clearly only the lower part of the curve could be
mapped out for the \bL phase because of the disorder in the
terminal ethylene groups.

\vspace{11pt}\noindent\textbf{Acknowledgements}\vspace{11pt}

\noindent I would like to thank Ross McKenzie and Jochen Wosnitza
for useful conversations. This work was supported by the
Australian Research Council.











\end{document}